# Non-Equilibrium Pathways for Excitation of Bulk and Surface Phonons through Anharmonic Coupling

C. Brand[1], V. Kümper[1], A. Hanisch-Blicharski[1,†], M. Tajik[1], J. D. Fortmann[1], A. Kaßen[1], F. Thiemann[1], M. Horn-von Hoegen[1,2,‡]

[1]*Department of Physics, University of Duisburg-Essen, 47057 Duisburg, Germany*

[2]*Center for Nanointegration (CENIDE), 47057 Duisburg, Germany*

[†]*present address: Institute of IT Management and Digitization Research (IFID), FOM University of Applied Sciences, 40476 Düsseldorf, Germany*

[‡]corresponding author: mhvh@uni-due.de

**Abstract**

Upon impulsive optical excitation of solid-state materials, the non-equilibrium flow of energy from the excited electronic system to the lattice degrees of freedom typically happens in a few picoseconds. Here we identified the surface of thin Bi films grown on Si(001) as an additional subsystem which is excited much slower on a 100 ps timescale that is caused by decoupling due to mismatched phonon dispersions relations of bulk and surface. Anharmonic coupling among the phonon systems provides pathways for excitations which exhibits a $1/T$-dependence causing a speed-up of surface excitation at higher temperatures. A quantitative justification is provided by phonon Umklapp processes from lattice thermal conductivity of the Bi bulk. Three-temperature model simulations reveal a pronounced non-equilibrium situation up to nanoseconds: initially, the surface is colder than the bulk, that situation is then inverted during cooling and the surface feeds energy back into the bulk phonon system.

The coupling of non-equilibrium excitations within the electron system to the nuclear position of atoms is ubiquitous in nature. It is a prerequisite for biological processes that govern life and is also integral to modern technology. We have only just begun to explore these dynamics for the rich world of solid-state materials on the relevant time and length scales. The use of coupled rate equations as in the two-temperature model was the starting point for describing the energy transfer between the hot electron system and the cold lattice by a single electron-phonon coupling constant.[1,2] More sophisticated models consider a larger number of phononic subsystems[3] and also include spin degrees of freedom.[4] The addition of *ab-inito* theory allows the description of the momentum- and energy-resolved coupling between the electron and the lattice system.[5,6]

A drosophila for such studies of out-of-equilibrium excitations triggered by ultrashort laser pulses is bismuth. As a Peierls-distorted semimetal with low carrier density in the electron and hole pockets,[7] Bi is very sensitive to optical excitations. Famous examples for structural dynamics observed in Bi include accelerated displacive excitation of the coherent $A_{1g}$ phonon mode,[8-14] mode softening and reversal of the Peierls distortion.[11,15] non-thermal melting.[15,16], and excitation of photoacoustic waves.[17,18]

The occurrence of this multitude of possible processes depends sensitively on the incident or absorbed laser fluence, i.e., density of excited carriers in the Bi. Strong excitation with fluences of more than 6 mJ/cm² causes a rapid and strong change in the potential energy surface resulting in nonthermal melting and destruction of the Bi film.[15,16] For lower fluences the lattice response is reversible, the coherent $A_{1g}$ phonon mode is launched,[19] and bond softening occurs[8,10,11,15,20-22] which ultimately results in an inverse Peierls transition.[23] Subsequently, the lattice is thermally heated on time scales of 2–4 ps[16,19,24-26] through energy transfer from the electron system to the lattice and coupling of the optical $A_{1g}$ phonon mode to acoustic phonons.[27]

All these experiments, except a study by Nagao *et al.*,[28] were performed *ex-situ* after handling the samples under ambient conditions, resulting in the formation of a stable passivating oxide layer with a thickness of ~3 nm.[29] Thus, all these studies provide solely information about the dynamics of the bulk and at the surface do not play any role.

Here we used ultrafast reflection high-energy electron diffraction (RHEED) [30] in an *in-situ* study under ultrahigh vacuum (UHV) conditions ($p = 3 \times 10^{-10}$ mbar) to investigate the surface lattice dynamics upon impulsive fs-laser excitation of ultrathin Bi films unaffected by an oxide layer as sketched in Fig. 1(a). Grazing incidence at 2.4° of the probing electrons accelerated to 30 keV ensures the surface sensitivity.[31,32] Using the Debye-Waller effect, the emergence of atomic motion $\Delta \mathbf{u}$ is directly accessible via the transient intensity suppression $I(\Delta t)/I_0 = \exp(-\langle \Delta \mathbf{u}\, \Delta \mathbf{k} \rangle^2)$ in the diffraction patterns with momentum transfer $\Delta \mathbf{k}$. We address the role of anharmonic coupling among phonon sub-systems to explain the delayed and strongly temperature-dependent excitation of thermal vibrational motion of surface atoms subsequent to the excitation of bulk phonons through conventional electron-phonon coupling.

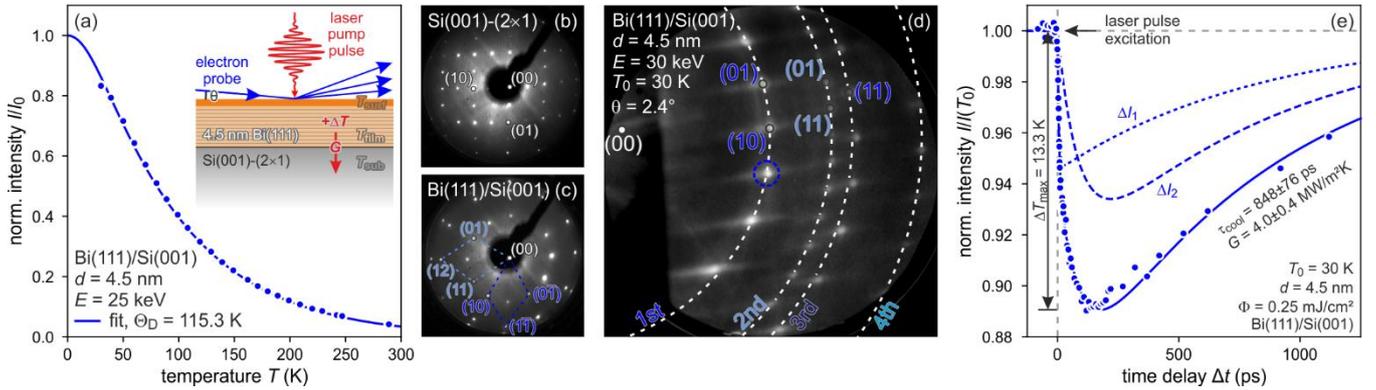

Fig. 1: (a) Debye-Waller analysis of the Bi diffraction spot marked by the blue circle in (d) as function of stationary sample temperature. Inset: sketch of Bi thin film sample, pump laser pulse, and probe electron pulse scheme. LEED patterns of (b) the bare Si(001)-(2×1) surface and (c) the Bi(111) film showing two by 90°-rotated single crystal domains. (d) RHEED pattern of the Bi(111) film. Laue circles are indicated by the dashed lines. (e) RHEED mean spot intensity from first Laue circle as function of time delay upon impulsive excitation at $\Delta t = 0$ with a fs-laser pulse.

The Si(001) sample was mounted on a liquid He-cooled cryostat providing sample temperatures 30 K < $T_0$ < 200 °C which were set by a built-in resistive heater. Heating to $T$ > 200 °C was facilitated through direct current passing through the Si sample.

A clean Si(001) surface was prepared by *in-situ* flash-annealing of the sample to $T$ > 1200 °C through thermal desorption of the native oxide. The LEED pattern in Fig. 1(b) shows the (2×1) reconstruction of the Si(001) substrate. The epitaxial Bi film was *in-situ* grown by molecular beam epitaxy following a recipe published by Jnawali *et al.*.[34] RHEED intensity oscillations during layer-by-layer growth of Bi[35] were used for calibration of the film thickness of $d = 4.5$ nm, i.e., 12 Bi bilayers with a spacing of $d_{Bi(111)} = 3.94$ Å.[7] Annealing of the Bi film at 200 °C eliminates surface roughness and provides a flat surface.[36,37] The LEED and RHEED diffraction patterns in Fig. 1(c,d) can be understood as incoherent superposition from the two by 90°-rotated Bi(111) domains on the (2×1)-reconstructed Si(001) surface with its inherent twofold symmetry. Thus, the first, third and fourth Laue circle of the RHEED pattern correspond to the same rotational domain, while the second Laue circle belongs to the other.

For the impulsive laser excitation of the Bi film, we used an amplified Ti:sapphire laser system with laser pulses of 50 fs length at a central wavelength of 800 nm and a pulse energy of 0.5 mJ at a repetition rate of 5 kHz. To record the transient changes of diffracted intensity, the time delay $\Delta t$ between pumping laser pulse at an incident fluence of 0.25 mJ/cm$^2$ and probing electron pulse was varied by an optomechanical delay line. The velocity mismatch of normal incidence pumping laser pulse and grazing incidence probing electron pulse was compensated by pulse front tilting by 71°.[38] Due to the high number of electrons in the probe pulse the overall temporal resolution of this experiment was limited to $\tau_{TIRF}$ ~ 6 ps.

The transient intensity decrease subsequent to impulsive optical excitation is shown in Fig. 1(e) for a sample temperature of $T_0 = 30$ K. A rapid drop of intensity by 11 % is followed by a slow exponential recovery with a time constant $\tau_{cool} = 848\pm76$ ps to the initial intensity prior to excitation. The maximum intensity change is reached at around $\Delta t$ ~ 175 ps. Assuming thermal equilibrium and utilizing the stationary Debye-Waller curve from Fig. 1(a) we thus determined a maximum temperature rise $\Delta T_{max} \cong +13$ K upon laser excitation which is small compared to the sample temperature. Thus, the experiment was performed in the regime of weak excitation. The same holds also for the other sample temperatures of our study.

The slow recovery to the initial intensity is determined through cooling of the Bi film via heat transfer across the Bi/Si interface towards the Si substrate. This process is limited by the interfacial thermal boundary conductance $G = c_V \rho_{Bi} d / \tau_{cool}$ with $\rho_{Bi}$ the mass density of Bi, and $c_V = 72.7$ J/kgK the specific heat[39] at the mean temperature between $T_0 = 30$ K and $T_{max} = T_0 + \Delta T_{max} = 43$ K. Here, we obtain a value $G = 4.0\pm0.4$ MW/m$^2$K which agrees with earlier studies,[36] thus confirming that the intensity drop is caused by the Debye-Waller effect and reflects the heating and subsequent cooling of the film.

The initial dynamics of the structural excitation were determined from the temporal fine structure of the intensity drop. For the lowest sample temperature of $T_0 = 30$ K this is shown in Fig. 1(e) for the first Laue circle (for the comparison with the second and third Laue circle and background intensity we refer to the Supplemental Material.). To improve the signal-to-noise ratio the intensities of all the spots on the Laue circle were averaged. The transient intensity exhibits a pronounced bi-exponential behavior as function of time delay $\Delta t$. First, we observe a rapid initial decrease of intensity with a time constant of $\tau_1$ ~ 5 ps. Second, there is a subsequent slower decrease of intensity with a time constant $\tau_2 = 94\pm17$ ps. We fitted this behavior with

$$\frac{I(\Delta t)}{I_0} = \left(1 - \Delta I_1 \cdot e^{-\Delta t/\tau_1} - \Delta I_2 \cdot e^{-\Delta t/\tau_2}\right) \cdot e^{-\Delta t/\tau_{cool}}, \quad (1)$$

for $\Delta t > 0$, with $\Delta I_1$ and $\Delta I_2$ the amplitudes of the fast and slow component [cf. dotted and dashed curves in Fig. 1(e)], respectively, and convoluted with the gaussian-shaped temporal response function of the instrument with full width at half maximum $\tau_{TIRF}$. The exponential with time constant $\tau_{cool}$ describes the cooling of the Bi film towards the Si substrate.

The changes of intensity of the spots on the first Laue circle (normalized to the intensity change at $\Delta t = 175$ ps for better visibility) for five different sample temperatures (30–300 K) are shown in Fig. 2. The bi-exponential behavior is obtained for all temperatures and was fitted with Eq. (1). We kept the ratio $\Delta I_1/\Delta I_2 = 0.57$ as a constant fitting parameter for the entire data set as will be justified later. The time constant $\tau_1$ ~ 5 ps of the rapid initial change of intensity appears independent on sample temperature $T_0$. In contrast, the time constant $\tau_2$ of the slower decrease of intensity depends strongly on $T_0$, ranging from 97±18 ps at 30 K to 12±30 ps at 300 K (Table S1 in the Supplemental Material). The cooling rate from the Bi film to the substrate slightly increases from lower to higher temperature since the corresponding time constant decreases from $\tau_{cool}$ = 848±76 ps at 30 K to 572±103 ps at 300 K.

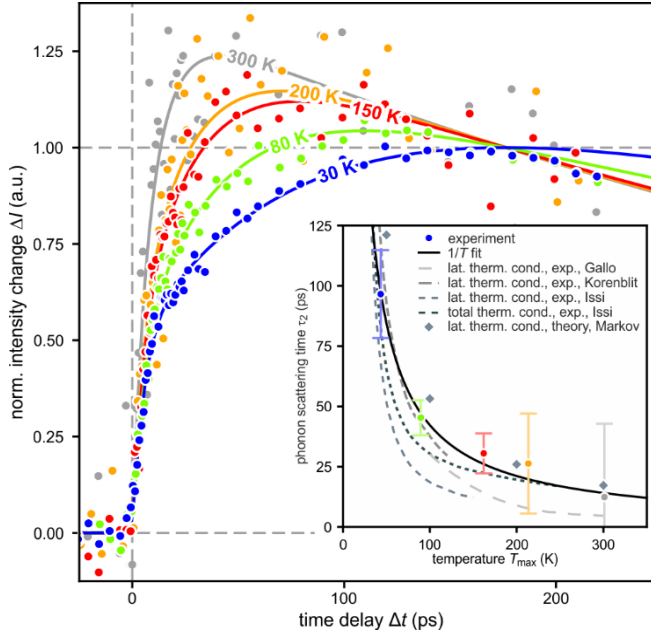

Fig. 2: Transient changes of intensity (normalized to $\Delta t$ = 175 ps) for different sample temperatures $T_0$. Solid lines give the fit to a bi-exponential behavior for the rise and to a mono-exponential decay for the decrease of the data points through cooling to the substrate. Inset: Time constants $\tau_2$ plotted as function of maximum temperature $T_{max}$. A $1/T$-behavior is fitted to the data points (solid black line). Data determined from lattice thermal conductivity measurements[40-42] and theory[43] are shown for comparison.

We expect an uniform excitation of the bulk for a 4.5 nm thick Bi film due to the almost homogeneous optical absorption of the laser pulse with an absorption length of 15 nm.[44-47] Ultrafast transport of the excited charge carriers on a 100 fs-timescale further leads to spatial homogeneity of the excitation.[10,48,49] Due to the weak excitation conditions (absorbed fluence in the Bi film of only 3–7 μJ/cm$^2$ in our study) we can safely exclude contributions from squeezed phonon states as it was reported by Johnson *et al*..[14] We therefore attribute the rapid initial drop of intensity to the excitation of lattice motion in the Bi bulk on a 2–4 ps-timescale [16,19,24-26] through electron-phonon coupling which renders this mechanism independent on sample temperature $T_0$.

The second contribution to the bi-exponential behavior exhibits a much slower decrease of intensity. Such a slow contribution has never been reported in other studies by time-resolved transmission electron diffraction or X-ray diffraction which both are sensitive only to the bulk of oxidized Bi samples. We therefore attribute this second contribution to the thermal excitation of the non-oxidized bare surface of the Bi film. This assumption is supported by the surface sensitivity of the diffraction geometry utilizing grazing incidence.

Even though the RHEED technique is surface sensitive, it still exhibits exponentially decaying sensitivity into the bulk.[31,32] Thus, we also observe contributions to the transient intensity change from the Bi bulk which justifies a fixed amplitude ratio $\Delta I_1/\Delta I_2$ of fast and slow component for all temperatures.

The pronounced temperature dependence of this second slow contribution is evidence against an excitation scenario mediated by electron-phonon coupling. Instead, we propose coupling from the excited bulk phonon modes to surface phonon modes.

Since the momentum-resolved phonon states of the surface do not overlap with those from the bulk,[33,50] the nuclear motion of the surface atoms can only be excited by anharmonic phonon-phonon processes.[51] These provide a more efficient coupling between bulk and surface phonons at higher sample temperatures when the atoms sample the non-parabolic regime of the potential energy landscape. Accordingly, this process is slowest at the lowest temperature of 30 K where it takes $\tau_2 = 97$ ps to heat the surface layer. At 300 K the vibrational amplitude is larger by a factor $(300\text{ K}/30\text{ K})^{1/2} \cong 3.2$ and accordingly the surface heats up much faster within $\tau_2 = 12$ ps.

For a quantitative explanation of the strong temperature dependence of coupling between bulk and surface phonons, we estimate the expected excitation time constants for surface heating from the thermal conductivity of the bulk. The lattice thermal conductivity $\kappa_{ph}$ at high temperatures is dominated by Umklapp scattering. Such 3-phonon interactions become only possible through anharmonic phonon-phonon coupling. The phonon scattering time $\tau_{ph} = 3\kappa_{ph}/c_V\rho_{Bi}v^2$ can thus be derived from the temperature dependencies of $\kappa_{ph}$ and the specific heat $c_V$. For simplicity, the mass density $\rho_{Bi}$ and the mean velocity of sound $v = 1288$ m/s of Bi[52] were treated as independent on temperature. The phonon scattering times in Bi determined from experimental[40-42] and theorical[43] results for $\kappa_{ph}$ are shown together with our experimentally derived values for $\tau_2$ in the inset of Fig. 2. For Bi, Umklapp scattering sets in at $T \approx 3$ K, well above the Casimir and dielectric range when $\kappa_{ph}$ is determined by the sample dimensions and the $T^3$-dependence of $c_V$, respectively. We therefore can safely assume a $1/T$-behavior for the lattice thermal conductivity[53] for $T > \Theta_D/4$ where $\Theta_D = 119$ K is the Bi bulk Debye temperature. A $1/T$-fit to our experimental data is shown as black solid line in the inset of Fig. 2. Despite the simplicity of our derivation, we find good agreement of our experimental time constants for excitation of the surface phonons in comparison with the phonon scattering times derived from literature.

To obtain further insight into the non-equilibrium energy flow and the associated coupling parameters we applied a three-temperature model to the surface-film-substrate system. We modelled the temperatures of the three subsystems involved, i.e., the excited charge carriers with $T_{el}$, and the bulk and surface phonons of the Bi film with $T_{bulk}$ and $T_{surf}$, respectively, through three rate equations:

$$\frac{\mathrm{d}}{\mathrm{d}\Delta t}T_{el} = \frac{1}{c_{el}}[A(\Delta t) - g_{el-b} \cdot (T_{el} - T_{bulk})], \quad (2a)$$

$$\frac{\mathrm{d}}{\mathrm{d}\Delta t}T_{surf} = \frac{N}{c_V}g_{b-s} \cdot (T_{bulk} - T_{surf}), \quad (2b)$$

$$\frac{\mathrm{d}}{\mathrm{d}\Delta t}T_{bulk} = \frac{N}{N-1} \cdot \frac{1}{c_V}[g_{el-b} \cdot (T_{el} - T_{bulk}) - g_{b-s} \cdot (T_{bulk} - T_{surf}) - g_{b-Si} \cdot (T_{bulk} - T_0)], \quad (2c)$$

considering the different spatial dimensions of the subsystems involved, i.e., with the $N = 12$ the number of bilayers of the 4.5 nm thin Bi film. For simplicity, we assumed the same temperature-dependent specific heat $c_V$ for the surface bilayer as for the Bi bulk. A Gaussian source term $A(\Delta t)$ describes the optical excitation of the electron system through the laser pulse.

The energy transfer between the three subsystems is described by two coupling parameters $g_{el-b}$ and $g_{b-s}$ for coupling between charge carriers to bulk phonons and bulk to surface phonons, respectively. Excitation from the charge carrier system to surface phonons was neglected, $g_{el-s} = 0$ due to the slow excitation of the surface.

Due to its high thermal conductivity the Si substrate was treated as efficient heat sink with constant temperature $T_0$.[36] The heat transfer from the Bi film to the Si substrate is determined by the thermal boundary conductance $G$ of the Bi bulk without the surface bilayer to the Si substrate and modelled through the coupling parameter $g_{b-Si} = G/(\rho_{Bi} \cdot (N-1) d_{Bi(111)}) = \rho_{Bi} \cdot c_v/\tau_{cool}$ (Table S1 in the Supplemental Material). The Schottky barrier of ~0.6 eV between Bi film and Si substrate partially hinders charge transfer across the interface.[54] For simplicity, we do not consider $g_{el-Si}$ in Eq. (2) because this loss of excited carriers into the substrate is already attributed to the source term $A(\Delta t)$.

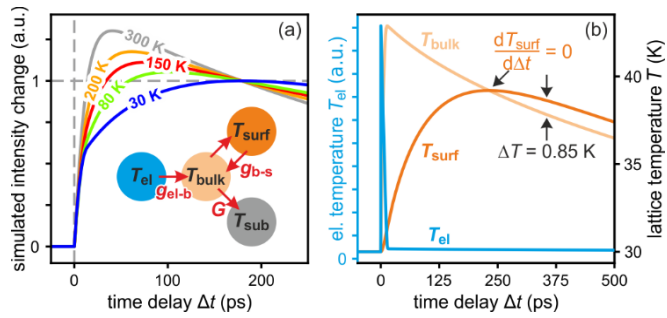

Fig. 4: Three-temperature model of (a) transient intensity for different sample temperatures $T_0$ and (b) evolution of temperatures of the three subsystems, i.e., excited charge carriers (blue), bulk (yellow) and surface phonons (orange) of the Bi film.

The coupling parameter $g_{el-b} \approx 1.6 \times 10^{15}$ W/m³K between excited carriers (electrons and holes) and bulk phonons is determined from the rise time of Bi bulk temperature on the order of 5 ps.

Now, the coupling parameter $g_{b-s}$ between bulk and surface phonons was determined by fitting the model curves to the experimental transient temperature rise $\Delta T$ for the individual base temperatures $T_0$ (Table S1 in the Supplemental Material). While the coupling between bulk and substrate only increases by a factor of ~2.5 in comparison of the lowest and the highest temperature, the coupling between bulk and surface increases by a factor of ~15, which corroborates our finding of anharmonic coupling of bulk and surface. Assuming a $1/T$-behavior for the anharmonic coupling and the same ratio for the sensitivity of the Debye-Waller effect to the displacements of bulk and surface phonons (as observed in the experiment) we were able to model the experimental behavior as depicted in Fig. 3(a). The evolution of temperatures for the three subsystems is plotted in Fig. 3(b) for $T_0 = 30$ K (individual curves for all values of $T_0$ are supplied in the Supplemental Material). While the bulk lattice system is heated on a few-ps timescale we observe delayed heating of the surface.

Interestingly, we observe a pronounced out-of-equilibrium situation between the bulk and surface lattice system. While on short timescales the surface is cooler than the bulk this situation is inverted on longer timescales such that the surface feeds back energy to the bulk. This inversion of temperature with $T_{surf} > T_{bulk}$ sets in when the surface temperature has reached its maximum. Cooling of the surface layer is delayed due to the additional barrier for heat transfer between surface and bulk. This non-equilibrium situation is present for more than nanoseconds until the bulk has reached the substrate temperature.

The bi-exponential decrease in diffraction intensify described in this study exhibits qualitatively different dynamics than those reported for thin Bi(111) films grown on Si(111) substrates[55] where a mono-exponential decrease of intensity has been observed without evidence for a fast component.[56] We attribute this surprising discrepancy to the different strain state of the epitaxially grown Bi(111) films on Si substrates with different orientations. On Si(111) the films grow in a relaxed manner with their bulk lattice parameter.[29,57] Bi films grown on Si(001) substrates, however, are uniaxially compressed by 2.3 % for film thicknesses below 6 nm.[58] Such a significant distortion of the lattice affects the electronic properties of the Bi film: not only the overlap of the bands may change but also the presence of the surface state. In this respect, Bi(111) films epitaxially grown on these two Si substrates are not identical and may in fact exhibit qualitatively different behavior upon impulsive optical excitation. Here we observed vibrational excitation of the surface through anharmonic coupling to bulk phonons.

Independent on details of the interpretation, we observed a long-lasting – more than many 100 ps – non-equilibrium between the bulk and the surface phonon system which is caused by the mismatch of their phonon dispersions. During its excitation the surface remains cold while the bulk of the Bi film is already hot. Over the long-lasting cooling, we observe the opposite: the surface is hotter than the Bi bulk for more than a nanosecond. Excitation and de-excitation of the surface layer are facilitated only through anharmonic coupling between the two phononic subsystems.

**Acknowledgements**

We gratefully acknowledge fruitful discussions with B. Rethfeld, W.G. Schmidt, U. Gerstmann, T. Biktagirov, S. Ito, I. Matsuda, P. Kratzer, and M. Gruner. Funded by the Deutsche Forschungsgemeinschaft (DFG, German Research Foundation) through Collaborative Research Center SFB 1242 "Non-equilibrium dynamics of condensed matter in the time domain" (Project-ID 278162697).

**References**

[1] S. I. Anisimov, B. L. Kapeliovich, and T. L. Perel'man,
*Electron emission from metal surfaces exposed to ultrashort laser pulses,*
Zh. Eksp. Teor. Fiz **66**, 776-781 (1974) [Sov. Phys. JETP **39,** 375–377 (1974)]

[2] B. Rethfeld, D. S. Ivanov, M. Garcia, and S. I. Anisimov,
*Modelling ultrafast laser ablation*,
J. Phys. D: Appl. Phys. **50,** 193001 (39pp) (2017)
doi: 10.1088/1361-6463/50/19/193001

[3] L. Waldecker, R. Bertoni, R. Ernstorfer, and J. Vorberger,
*Electron-Phonon Coupling and Energy Flow in a Simple Metal beyond the Two-Temperature Approximation,*
Phys. Rev. X **6**, 021003 (2016)
doi: 10.1103/PhysRevX.6.021003

[4] D. Zahn, F. Jakobs, Y. W. Windsor, H. Seiler, T. Vasileiadis, T. A. Butcher, Y. Qi, D. Engel, U. Atxitia, J. Vorberger, and R. Ernstorfer,
*Lattice dynamics and ultrafast energy flow between electrons, spins, and phonons in a 3d ferromagnet,*
Phys. Rev. Research **3**, 023032 (2021)
doi: 10.1103/PhysRevResearch.3.023032

[5] L. Waldecker, R. Bertoni, H. Hübener, T. Brumme, T. Vasileiadis, D. Zahn, A. Rubio, and R. Ernstorfer,
*A Momentum-Resolved View on Electron-Phonon Coupling in Multilayer WSe2*,
Phys. Rev. Lett. **119**, 036803 (2017)
doi: 10.1103/PhysRevLett.119.036803

[6] P. Kratzer, L. Rettig, I. Yu. Sklyadneva, E. V. Chulkov, and U. Bovensiepen,
*Relaxation of photoexcited hot carriers beyond multitemperature models: General theory description verified by experiments on Pb/Si(111),*
Phys. Rev. Res. **4**, 033218 (2022)

[7] Ph. Hofmann
*The surfaces of bismuth: Structural and electronic properties,*
Progress in Surface Science **81**, 191–245 (2006)
doi: 10.1016/j.progsurf.2006.03.001

[8] M. Hase, K. Mizoguchi, H. Harima, S. Nakashima, and K. Sakai,
*Dynamics of coherent phonons in bismuth generated by ultrashort laser pulses,*
Phys. Rev. B **58**, 5448 (1998)
doi: 10.1103/PhysRevB.58.5448

[9] A. A. Melnikov, O. V. Misochko, and S. V. Chekalin
*Generation of coherent phonons in bismuth by ultrashort laser pulses in the visible and NIR: Displacive versus impulsive excitation mechanism,*
Physics Letters A **375**, 2017–2022 (2011)
doi: 10.1016/j.physleta.2011.03.057

[10] S. L. Johnson, P. Beaud, C. J. Milne, F. S. Krasniqi, E. S. Zijlstra, M. E. Garcia, M. Kaiser, D. Grolimund, R. Abela, and G. Ingold,
*Nanoscale Depth-Resolved Coherent Femtosecond Motion in Laser-Excited Bismuth.*
Phys. Rev. Lett. **100**, 155501 (2008)
doi: 10.1103/PhysRevLett.100.155501

[11]  S. W. Teitelbaum, T. Shin, J. W. Wolfson, Y-H. Cheng, I. J. Porter, M. Kandyla, and K. A. Nelson,
*Real-Time Observation of a Coherent Lattice Transformation into a High-Symmetry Phase*, Physical Review X **8**, 031081 (2018)
doi: 10.1103/PhysRevX.8.031081

[12]  A. Bugayev and H. E. Elsayed-Ali
*Lattice dynamics and electronic Grüneisen parameters of femtosecond laser-excited bismuth,*
Journal of Physics and Chemistry of Solids **129**, 312 (2019)
doi: 10.1016/j.jpcs.2019.01.030

[13]  R. Géneaux, I. Timrov, C. J. Kaplan, A. D. Ross, P. M. Kraus, and S. R. Leone
*Coherent energy exchange between carriers and phonons in Peierls-distorted bismuth unveiled by broadband XUV pulses*,
Phys. Rev. Research **3**, 033210 (2021)
doi: 10.1103/PhysRevResearch.3.033210

[14]  S. L. Johnson, P. Beaud, E. Vorobeva, C. J. Milne, E. D. Murray, S. Fahy, and G. Ingold,
*Directly observing squeezed phonon states with femtosecond x-ray diffraction,*
Phys. Rev. Lett. **102**, 175503 (2009)
doi: 10.1103/PhysRevLett.102.175503

[15]  D. M. Fritz *et al*.
*Ultrafast Bond Softening in Bismuth: Mapping a Solid`s Interatomic Potential with X-rays,*
Science **315**, 633–636 (2007)
doi: 10.1126/science.1135009

[16]  G. Sciani, M. Harb, S. G. Kruglik, T. Payer, C. T. Hebeisen, F.-J. Meyer zu Heringdorf, M. Yamaguchi, M. Horn-von Hoegen, R. Ernstorfer, and R. J. D. Miller,
*Electronic acceleration of atomic motions and disordering in bismuth,*
Nature **458**, 56–59 (2009)
doi: 10.1038/nature07788

[17]  T. Shin,
*Femtosecond reflectivity study of photoacoustic responses in bismuth thin films,*
Thin Solid Films, **666**, 108 (2018)
doi: 10.106/j.tsf.2018.09.037

[18]  A. Kumar, G. Sachdeva, R. Pandey, and S. P. Karna,
*Optical absorbance in multilayer two-dimensional materials: Graphene and antimonene,*
Appl. Phys. Lett. **116**, 263102 (2020)
doi: 10.1063/5.0010794

[19]  K. Sokolowski-Tinten, C. Blome, J. Blums, A. Cavalleri, C. Dietrich, A. Tarasevitch, I. Uschmann, E. Forster, M. Kammler, M. Horn-von-Hoegen, and D. Linde,
*Femtosecond X-ray measurement of coherent lattice vibrations near the Lindemann stability limit,*
Nature (London) **422**, 287–289 (2003)
doi: 10.1038/nature01490

[20]  M. Hase, M. Kitajima, S.-I. Nakashima, and K. Mizoguchi
*Dynamics of Coherent Anharmonic Phonons in Bismuth Using High Density Photoexcitation*,
Phys. Rev. Lett. **88**, 067401 (2002)
doi: 10.1103/PhysRevLett.88.067401

[21]  M. F. DeCamp, D. A. Reis, P. H. Bucksbaum, and R. Merlin,
*Dynamics and coherent control of high-amplitude optical phonons in bismuth*,
Phys. Rev. B **64**, 092301 (2001)
doi: 10.1103/PhysRevB.64.092301

[22]  E. D. Murray, D. M. Fritz, J. K. Wahlstrand, S. Fahy, D. A. Reis,
*Effect of lattice anharmonicity on high-amplitude phonon dynamics in photoexcited bismuth*,
Phys. Rev. B **72**, 060301(R) (2005)
doi: 10.1103/PhysRevB.72.060301

[23]  K. Sokolowski-Tinten,
*private communication*

[24]  A. R. Esmail and H. E. Elsayed-Ali,
*Anisotropic response of nanosized bismuth films upon femtosecond laser excitation monitored by ultrafast electron diffraction,*
Appl. Phys. Lett. **99**, 161905 (2011)
doi: 10.1063/1.3652919

[25]  C. Streubühr, A. Kalus, P. Zhou, M. Ligges, A. Hanisch-Blicharski, M. Kammler, U. Bovensiepen, M. Horn-von Hoegen, and D. von der Linde,
*Comparing ultrafast surface and bulk heating using time-resolved electron diffraction,*
Appl. Phys. Lett. **104**, 161611 (2014)
doi: 10.1063/1.4872055

[26]  K. Sokolowski-Tinten, R. Li, A. Reid, S. Weathersby, F. Quirin, T. Chase, R. Coffee, J. Corbett, A. Fry, N. Hartmann, C. Hast, R. Hettel, M. Horn-von Hoegen, D. Janoschka, J. Lewandowski, M. Ligges, F. J. Meyer zu Heringdorf, X. Shen, T. Vecchione, C. Witt, J. Wu, H. Dürr, and X. Wang,
*Thickness-dependent electron-lattice equilibration in laser-excited thin bismuth films,*
New J. Phys. **17**, 113047 (2015)
doi: 10.1088/1367-2630/17/11/113047

[27]  H. Hase, K. Ishioka, M. Kitajima, S. Hishita, and K. Ushida,
*Dephasing of coherent THz phonons in bismuth studied by femtosecond pump probe technique*,
Appl. Surf. Sci. **197–198**, 710–714 (2002)
doi: 10.1016/S0169-4332(02)00398-7

[28]  K. Ishioka, M. Kitajima, O. V. Misochko, and T. Nagao
*Ultrafast phonon dynamics of epitaxial atomic layers of Bi on Si(111)*,
Phys. Rev. B **91**, 125431 (2015)
doi: 10.1103/PhysRevB.91.125431

[29]  T. Payer, C. Klein, M. Acet, V. Ney, M. Kammler, F.-J. Meyer zu Heringdorf, and M. Horn-von Hoegen,
*High-quality epitaxial Bi(111) films on Si(111) by isochronal annealing,*
Thin Solid Films **520**, 6905–6908 (2012)
doi: 10.1016/j.tsf.2012.06.004

[30] A. Janzen, B. Krenzer, O. Heinz, P. Zhou, D. Thien, A. Hanisch, F. J. Meyer zu Heringdorf, D. von der Linde, and M. Horn-von Hoegen,
*A pulsed electron gun for ultrafast electron diffraction at surfaces*,
Rev. Sci. Instrum. **78**, 013906 (2007)
doi: 10.1063/1.2431088

[31] W. Braun,
*Applied RHEED: Reflection High-Energy Electron Diffraction During Crystal Growth*,
Springer, Berlin, Heidelberg (1999)
doi: 10.1007/BFb0109548

[32] A. Ichimiya and P. I. Cohen,
*Reflection High Energy Electron Diffraction*,
Cambridge University Press, Cambridge, UK (2004)
doi: 10.1017/CBO9780511735097

[33] G. Q. Huang and J. Yang,
*Surface lattice dynamics and electron–phonon interaction in ultrathin Bi(111) film*,
J. Phys.: Condens. Matter **25**, 175004 (2013)
doi: 10.1088/0953-8984/25/17/175004

[34] G. Jnawali, H. Hattab, B. Krenzer, and M. Horn von Hoegen,
*Lattice accommodation of epitaxial Bi(111) films on Si(001) studied with SPA-LEED and AFM*,
Phys. Rev. B **74**, 195340 (2006)
doi: 10.1103/PhysRevB.74.195340

[35] G. Jnawali, H. Hattab, C. Bobisch, A. Bernhart, E. Zubkov, R. Möller, and M. Horn-von Hoegen,
*Homoepitaxial growth of Bi(111)*,
Phys. Rev. B **78**, 035321 (2008)
doi: 10.1103/PhysRevB.78.035321

[36] A. Hanisch-Blicharski, V. Tinnemann, S. Wall, F. Thiemann, Th. Groven, J. Fortmann, M. Tajik, C. Brand, B.O. Frost, A. von Hoegen, and M. Horn-von Hoegen,
*Violation of Boltzmann Equipartition Theorem in Angular Phonon Phase Space Slows down Nanoscale Heat Transfer in Ultrathin Heterofilms*,
Nano Lett. **21**, 7145–7151 (2021)
doi: 10.1021/acs.nanolett.1c01665

[37] C. Bobisch, A. Bannani, M. Matena and R. Möller,
*Ultrathin Bi films on Si(100)*,
Nanotechnology **18**, 055606 (2007)
doi: 10.1088/0957-4484/18/5/055606

[38] P. Zhou, C. Streubühr, A. Kalus, T. Frigge, S. Wall, A. Hanisch-Blicharski, M. Kammler, M. Ligges, U. Bovensiepen, D. von der Linde, and M. Horn-von Hoegen,
*Ultrafast time resolved reflection high energy electron diffraction with tilted pump pulse fronts*,
EPJ Web Conf. **41**, 10016 (2013)
doi: 10.1051/epjconf/20134110016

[39]     L. G. Carpenter, T. F. Harle, and A. C. Egerton,
        *The atomic heat of bismuth at higher temperatures*,
        Proceedings of the Royal Society of London. Series A, Containing Papers of a Mathematical and Physical Character **136**, 243–250 (1932)
        doi: 10.1098/rspa.1932.0077

[40]     C. F. Gallo, B. S. Chandrasekhar, and P. H. Sutter,
        *Transport Properties of Bismuth Single Crystals*,
        J. Appl. Phys. **34**, 144–152 (1963)
        doi: 10.1063/1.1729056

[41]     I. Ya. Korenblit, M. E. Kuznetsov, V. M. Muzhdaba, and S. S. Shalyt,
        *Electron Heat Conductivity and the Wiedemann-Franz Law for Bi*,
        Zh. Eksp. Teor. Fiz. **57**, 1867–1876 (1969) [Sov. Phys. JETP **30**, 1009–1014 (1970)]

[42]     J.-P. Issi,
        *Low temperature transport properties of the group V semimetals*,
        Australian Journal of Physics **32**, 585 (1979)
        doi: 10.1071/PH790585

[43]     M. Markov,
        Ph.D. thesis,
        Université Paris Saclay, 2016.
        https://pastel.hal.science/tel-01438827

[44]     R. Scholz, T. Pfeifer and H. Kurz,
        *Density-matrix theory of coherent phonon oscillations in germanium*,
        Phys. Rev. B **47**, 16229 (1993)
        doi: 10.1103/PhysRevB.47.16229

[45]     D. Boschetto, E. G. Gamaly, A. V. Rode, B. Luther-Davies, D. Glijer, T. Garl, O. Albert, A. Rousse, J. Etchepare,
        *Small Atomic Displacements Recorded in Bismuth by the Optical Reflectivity of Femtosecond Laser-Pulse Excitations,*
        Phys. Rev. Lett. **100**, 027404 (2008)
        doi: 10.1103/PhysRevLett.100.027404

[46]     O. Hunderi,
        *Opitcal properties of crystalline and amorphous bismuth films,*
        J. Phys. F: Met. Phys. **5**, 2214–2225 (1975)
        doi: 10.1088/0305-4608/5/11/034/meta

[47]     R. Serna, J. Toudert, J. Gonzalo, A. Mariscal, E. Soria, and P. Gomez-Rodriguez,
        *Oxide-Based Luminiscent and Active Nanophotonic Structures,*
        Meet. Abstr. **MA2020-01**, 1083 (2020)
        doi: 10.1149/MA2020-01161083mtgabs/meta

[48]     G. Jnawali, D. Boschetto, L. M. Malard, T. F. Heinz, G. Sciaini, F. Thiemann, T. Payer, L. Kremeyer, F.-J. Meyer zu Heringdorf, and M. Horn-von Hoegen,
        *Hot carrier transport limits the displacive excitation of coherent phonons in bismuth,*
        Appl. Phys. Lett. **119**, 091601 (2021)
        doi: 10.1063/5.0056813

[49] F. Thiemann, G. Sciaini, A. Kassen, U. Hagemann, F. Meyer zu Heringdorf, and M. Horn-von Hoegen, *Ultrafast transport mediated homogenization of photoexcited electrons governs the softening of the A1g phonon in bismuth,*
Phys. Rev. B **106**, 014315 (2022)
doi: 10.1103/PhysRevB.106.014315

[50] M. Alcántara Ortigoza, I. Y. Sklyadneva, R. Heid, E. V. Chulkov, T. S. Rahman, K.-P. Bohnen, and P. M. Echenique,
*Ab initio lattice dynamics and electron-phonon coupling of Bi(111),*
Phys. Rev. B **90**, 195438 (2014)
doi: 10.1103/PhysRevB.90.195438

[51] S. Sakong, P. Kratzer, S. Wall, A. Kalus, and M. Horn-von Hoegen,
*Mode conversion and long-lived vibrational modes in lead monolayers on silicon (111) after femtosecond laser excitation: A molecular dynamics simulation,*
Phys. Rev. B **88**, 115419 (2013)
doi: 10.1103/PhysRevB.88.115419

[52] Y. Eckstein, A. W. Lawson, and D. H. Reneker, *Elastic Constants of Bismuth*,
J. Appl. Phys. **31**, 1534–1538 (1960)
doi: 10.1063/1.1735888

[53] K. W. Böer and U. W. Pohl, *Semiconductor Physics,*
Springer International Publishing, 2020
doi: 10.1007/978-3-319-06540-3

[54] K. Hricovini, G. Le Lay, A. Kahn, A. Taleb-Ibrahimi, and J. E. Bonnet,
*Initial stages of Schottky-barrier formation of Bi/Si(111) and Bi/Si(100) interfaces,*
Appl. Surf. Sci. **56–58**, 259–263 (1992)
doi: 10.1016/0169-4332(92)90244-R

[55] V. Tinnemann, C. Streubühr, B. Hafke, A. Kalus, A. Hanisch-Blicharski, M. Ligges, P. Zhou, D. von der Linde, U. Bovensiepen, and M. Horn-von Hoegen
*Ultrafast electron diffraction from a Bi(111) surface: Impulsive lattice excitation and Debye–Waller analysis at large momentum transfer,*
Struct. Dyn. **6**, 035101 (2019)
doi: 10.1063/1.5093637

[56] V. Tinnemann, C. Streubühr, B. Hafke, T. Witte, A. Kalus, A. Hanisch-Blicharski, M. Ligges, P. Zhou, D. von der Linde, U. Bovensiepen, and M. Horn-von Hoegen,
*Decelerated vibrational excitation and absence of bulk phonon modes at surfaces: ultrafast electron diffraction from Bi(111) surface upon fs-laser excitation,*
Struct. Dyn. **6**, 065101 (2019)
doi: 10.1063/1.5128275

[57] M. Kammler and M. Horn-von Hoegen,
*Low energy electron diffraction of epitaxial growth of bismuth on Si(111),*
Surf. Sci. **576**, 56–60 (2005)
doi: 10.1016/j.susc.2004.11.033

[58] D. Meyer, G. Jnawali, H. Hattab, and M. Horn-von Hoegen,
*Rapid Onset of Strain Relief by Mass Generation of Misfit Dislocations in Bi(111)/Si(001) Heteroepitaxy*,
Appl. Phys. Lett. **114**, 081601 (2019)
doi: 10.1063/1.5088760

## Supplemental Material

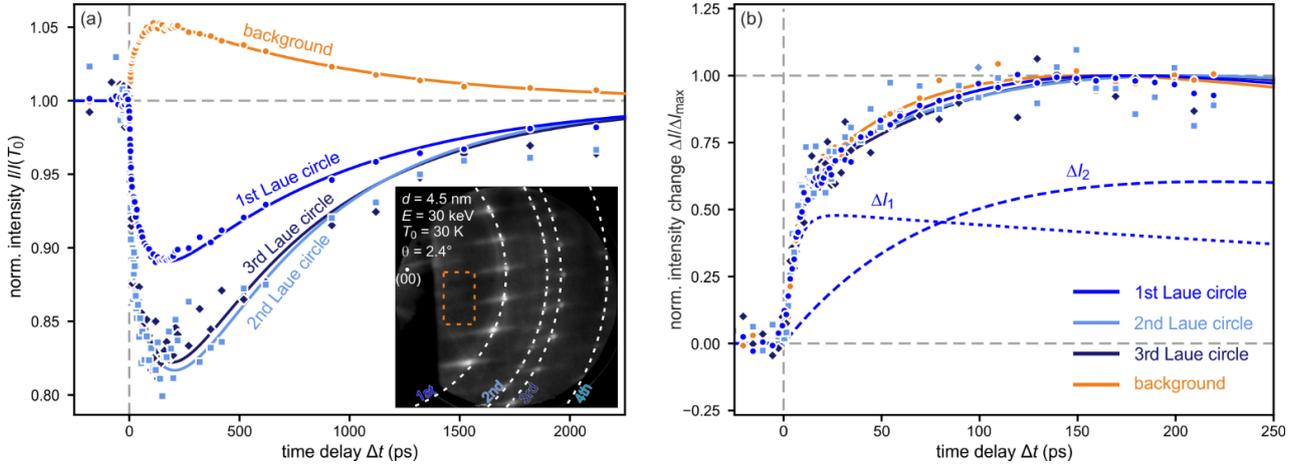

Fig. S1: (a) Transient RHEED intensity upon impulsive excitation at a sample temperature of $T_0 = 30$ K. The data were averaged over the spots of the three individual Laue circles and the diffuse background intensity taken from the area indicated by the orange rectangle in the RHEED pattern (inset). Solid lines give fits according to Eq. (1). (b) Transient intensity changes normalized to the maximum value $\Delta I_{max}$ for the three Laue circles and background. The two contributions $\Delta I_1$ and $\Delta I_2$ giving rise to the fast and the slow intensity drops of the first Laue circle are shown by the dotted and dashed lines, respectively.

The initial dynamics of the structural excitation were determined from the temporal fine structure of the intensity drop, while the recovery of intensity is mediated through cooling of the Bi bulk to the underlying Si substrate. For the lowest sample temperature of $T_0 = 30$ K both is shown in Fig. S1 for the first three Laue circles and the background intensity. To improve the signal-to-noise ratio the intensities of all the spots on each of the Laue circles were averaged. Hereby, the first (L1) and third Laue circle (L3) belong to one rotational domain of the Bi(111) film grown on Si(001), while the second (L2) and fourth Laue circle (L4) originate from the by 90° rotated domain.

The transient intensities for all three Laue circles exhibit a pronounced bi-exponential behavior as function of time delay $\Delta t$. The rapid initial decrease of intensity occurs at a time constant of $\tau_1 \sim 5$ ps, independent on the order of the Laue circle. The increase of diffuse background intensity shows the same temporal behavior. For the slow excitation time constant we measured values of $\tau_{2,L1} = 94\pm17$ ps, $\tau_{2,L2} = 191\pm201$ ps, and $\tau_{2,L3} = 138\pm71$ ps for the first three Laue circles and $\tau_{2,b} = 78\pm21$ ps for the background.

The mono-exponential recovery of intensity with time constant $\tau_{cool}$ describes the cooling of the Bi film towards the Si substrate. The cooling time constants were determined as $\tau_{cool,L1} = 857\pm76$ ps, $\tau_{cool,L2} = 690\pm326$ ps, and $\tau_{cool,L3} = 758\pm192$ ps for the first three Laue circles and $\tau_{cool,b} = 834\pm95$ ps for the background.

Within the margins of error we do not observe any systematic deviations for the three time constants or for the intensity ratio $\Delta I_1/\Delta I_2 = 0.57$ of the fast and slow contribution as function of order of Laue circle as shown for the normalized intensity change $\Delta I/\Delta I_{max}$ in Fig. 1(b).

The maximum intensity change is reached at $\Delta t \approx 175$ ps. The same temporal response is observed for all spots and the diffuse background intensity thus supporting the interpretation as Debye-Waller behavior.

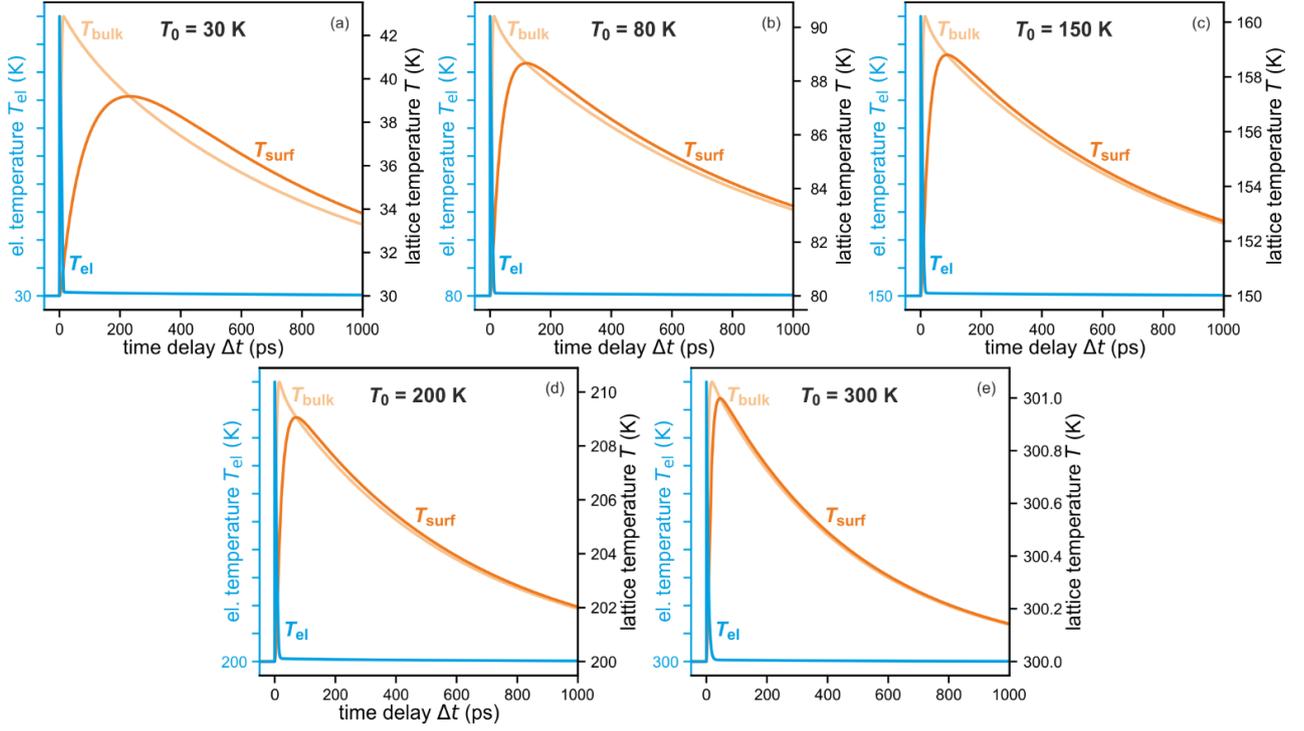

Fig. S2: Simulations for the electronic $T_{el}$ (blue), surface $T_{surf}$ (orange), and bulk $T_{bulk}$ (orange) temperatures for various base temperatures $T_0$ using three-temperature modelling.

The electronic $T_{el}$, surface $T_{surf}$, and bulk $T_{bulk}$ temperatures of a 12 bilayer thin Bi(111) film were computed according to Eq. (2a-c) for sample temperatures in the range of 15 K to 300 K (see Fig. S2). The coupling parameters (except for 15 K) were determined by fitting the simulated curves to the experimentally determined time constants for each sample temperature $T_0$.

For simplification, we assumed that the coupling parameters $g_{b-Si}$ and $g_{b-s}$ do not vary during the course of the transient temperature rise by $\Delta T_{max}$ for each sample temperature $T_0$ upon fs-laser excitation since we applied weak excitation densities.

| $T_0$ (K) | 30 | 80 | 150 | 200 | 300 |
|---|---|---|---|---|---|
| $\tau_2$ (ps) | 97±18 | 45±7 | 31±8 | 26±21 | 12±30 |
| $\tau_{cool}$ (ps) | 848±76 | 757±44 | 606±46 | 585±114 | 572±103 |
| $G$ (MW/m²K) | 4.0±0.4 | 6.6±0.4 | 8.8±0.7 | 9.2±1.8 | 9.5±1.7 |
| $\Delta T_{max}$ (K) | 13 K | 9 K | 12 K | 13 K | 1 K |
| $g_{b-s}$ ($10^{15}$ W/m³K) | 0.59 | 2.4 | 3.8 | 4.8 | 9.0 |
| $g_{b-Si}$ ($10^{15}$ W/m³K) | 0.84 | 1.4 | 1.9 | 2.0 | 2.0 |
| $c_V(T_0)$ (J/kgK) | 61 | 107 | 116 | 117 | 118 |

Table S1: Experimental values for time constants, thermal boundary conductance $G$, coupling constants $g_{b-s}$ and $g_{b-Si}$, and specific heat $c_V$ at $T_0$ for the 4.5 nm thin Bi(111) film grown on Si(001).